厦门大学

信息检索课程报告

# 基于内容的多媒体信息检索技术

王翌（31520091152835）

指导教师姓名：曹东林

专　业　名　称：计算机技术

2010 年 12 月 6 日

# 基于内容的多媒体信息检索技术

摘要：基于内容检索（Content-based Retrieval，简称CBR）实际指的是基于内容特征的检索。具体说来,是对媒体对象的内容及上下文语义环境所进行的检索。通俗的说，就是从媒体数据中提取出特定的信息线索，然后根据这些线索从大量存储在数据库中的媒体中进行查找，检索出具有相似特征的媒体数据出来。这类检索根据用户的要求,对文本、声音、图形、图像、动画等多媒体信息进行检索。基于内容的检索系统既能对以文本信息为代表的离散媒体进行检索,也能对以图像、声音为代表的连续媒体的内容进行检索。基于内容的多媒体信息检索包括图像检索，动态视频检索和音频检索等。本文介绍了基于内容的图像检索和基于内容的音频检索的部分内容。

关键词：基于内容检索，基于内容的图像检索，基于内容的音频检索。
.

# Content-based Multi-media Retrieval technology

**Abstract:** This paper gives a summary of the content-based Image Retrieval and Content-based Audio Retrieval, which are two parts of the Content-based Retrieval. Content-based Retrieval is the retrieval based on the features of the content. Generally, it is a way to extract features of the media data and find other data with the similar features from the database automatically. Content-based Retrieval can not only work on discrete media like texts, but also can be used on continuous media, such as video and audio.

# 目 录

# 1 前言：

  基于内容检索（Content-based Retrieval，简称CBR）实际指的是基于内容特征的检索。具体说来，是对媒体对象的内容及上下文语义环境所进行的检索。通俗的说，就是从媒体数据中提取出特定的信息线索，然后根据这些线索从大量存储在数据库中的媒体中进行查找，检索出具有相似特征的媒体数据出来。这类检索根据用户的要求,对文本、声音、图形、图像、动画等多媒体信息进行检索。基于内容的检索系统既能对以文本信息为代表的离散媒体进行检索,也能对以图像、声音为代表的连续媒体的内容进行检索。基于内容的多媒体信息检索包括图像检索，动态视频检索和音频检索等。

# 2 基于内容的图像与视频检索：

  如今，越来越多的图像和视频信息是以数字方式存储的，如何有效的管理和利用这些以图像和视频的方式存储的信息成为了一项十分重要的问题。

  目前在图像数据库和视频信息系统中，检索主要是根据随图像和视频信息一起存储的文本描述进行的，通过这些文本十分简洁，主要含有拍摄时间、地点、拍摄者信息，对图像本身的内容描述如果不是没有的话，也是十分简短的，难以满足实际检索时多方面的需要。基于内容的图像和视频检索研究的目的是直接根据图像和视频本身的信息，抽取检索特征，建立索引树，再根据一定的相似性衡量标准，实现检索。

  对于支持基于内容的图像检索的数据库而言，必然支持两种主要的数据结构：物体(objects)和场景(scenes)。其中场景是一副图像，其中可以含有一个或多个物体，也可以没有任何物体，而物体时场景的一部分，例如飞机是机场场景的一个物体。物体和场景这两种数据类型需要按照它们的视觉特征来代表。常用的图像特征包括颜色、纹理(texture)、形状(shape)、位置以及作为线条图草图(line sketches)时的边界之间的关系

  目前主要根据彩色直方图特征和纹理度量特征进行图像与视频检索。比较简单的方法有基于主颜色和颜色直方图的检索；而比较复杂的方法需要结合区域的位置

关系和几何形状信息按照分层的方法综合起来；同时，还可以根据用户的反馈信息利用机器学习的方法改进基于内容的图像检索算法；最终比较相似性的方法主要有欧氏距离和马氏距离等方法。

视频内容的标注与检索时十分复杂而困难的问题，目前与图像检索一样都处于初期阶段，在视频的标注方面，比较成熟的技术是镜头(shot)分割、关键帧(key frame)提取，在此基础上形成镜头组织，包括场景层次图(SHG, Scene Hierarchy Graph)和场景转换图(Scene Transition Graph)。但目前自动进行镜头组织的方法还很难达到令人满意的结果，人工交互组织不可避免。

## 3 基于内容的音频检索

经典的 IR 问题是利用一组关键字组成的查询来定位需要的文本文档，即定位文档中的查找关键字来发现匹配的文档。如果一个文档中包含较多的查询项，那么，它就被认为比其他包含较少查询项的文档更"相关"。于是，文档可以按照"相关度"来排序，并显示给用户，以便进一步搜索。但是，如果我们把数字音频当成一种不透明的位流来管理，虽然可以赋予名字、文件格式、采样率等属性，但是其中没有可以确认的词或者可以比较的实体。因此，不能像文本那样搜索活检索其内部的内容。对于音乐和非语音声响也是这样。

最先被提出的方法是基于人工输入的属性和描述来进行音频检索。该方法的主要缺点反应在以下几个方面：

1. 当数据量越来越多时，人工的注释强度加大
2. 人堆音频的感知，如音乐的旋律、音调、音质等，难以用文字注释表达清楚。

### 3.1 查询方法

音频是声音信号的形式。作为一种信息载体，音频可以分为三种类型：

1. 波形声音：波形声音是对模拟声音数字化而得到的数字音频信号，它可以代表语音、音乐、自然界和合成的声响；
2. 语音：语音具有自此、语法等语素，是一种高度抽象的概念交流媒体。语音

经过识别可以转换为文本。文本时语音的一种脚本形式。

3. 音乐：音乐具有节奏、旋律活和声等要素，是人声或/和乐器音响等配合所构成的一种声音。音乐可以用乐谱来表达。

不同的类型将具有不同的内在内容。但从整体上来看，音频内容分为三个级别：最低层的物理样本级、中间层的声学特征和最高层的语义级。

在物理样本级，音频内容呈现的是流媒体形式，用户可以通过实践刻度，检索活调用音频的样本数据。如现在常见的音频录放程序接口。

中间层是声学特征级。声学特征是从音频数据中自动抽取的。一些听觉特征表达用户对音频的感知，可以直接用于检测；一些特征用于语音的识别或检测，支持更高层的内容表示。另外还有音频的时空结构。

最高层是语义层，是音频内容、音频对象的概念级描述。具体来说，在这个级别上，音频的内容是语音识别、检测、辨别的结果，音乐旋律和叙事的说明，以及音频对象和概念的描述。

后两层是基于内容的音频检索技术最关心的。在这两个层次上，用户可以提交概念查询或按照听觉感知来查询。

音频的听觉特性决定其查询方式不同于常规的信息检索系统。基于内容的查询是一种相似查询，它实际上是检索出与用户指定的要求非常相似的所有声音。查询中可以指定返回的声音数或相似度的大小。另外，可以强调或关闭(忽略)某些特征成分，甚至可以施加逻辑"非"(或模糊的less匹配关系)来指定检索条件，检索那些不具有或少有某种特征成分(如指定没有"尖锐"或少有"尖锐")的声音。另外，还可以对给定的一组声音，按照声学特征进行排序，如按声音的嘈杂程度排序。

在查询接口上，用户可以采用以下形式提交查询：

1. 示例：用户可以采用以下形式提交查询该声音在某些特征方面相似的所有声音。
2. 直喻：用户通过选择一些声学/感知物理特征来描述查询要求，如亮度、音调、音量等。这种方式与可查询中的描绘查询方式。
3. 拟声：发出与要查找的声音兴致相似的声音来表达查询要求。如可以发出嗡嗡声来查找蜜蜂活电器嘈杂声。

4. 主观特征：用个人的描述语言来描述声音，这需要训练系统理解这些描述语的含义，如用户可能要寻找"欢快"的声音。
5. 浏览：这是信息发现的一种重要手段，尤其是对于音频这种时基媒体。除了在分类的基础上浏览目录外，重要的是基于音频的结构进行浏览。

## 3.2 音频检索技术

根据对音频媒体的划分可以知道,语音、音乐和其他声响具有显著不同的特性，因而目前的处理方法可以分为相应的三种：处理包含语音的音频和不包含语音的音频，后者又把音乐单独划分出来。换句话说，第一种是利用自动语音识别技术，后两种是利用更一般性的音频分析，以适合更广泛的音频媒体，如音乐和声音效果，当然也包含数字化语音信号。音频信息检索分为以下几方面：

### 3.2.1 基于语音技术的检索

语音检索是以语音为中心的检索，采用语音识别等处理技术。如电台节目、电话交谈、会议录音等。

基于语音技术的检索是利用语音处理技术检索音频信息。过去人们对语音信号处理开展了大量的研究，许多成果可以用于语音检索。

利用大词汇语音识别技术进行检索：利用自动予以能识别(ASR)技术把语音转换为文本，从而可以采用文本检索方法进行检索。虽然好的连续语音识别系统在小心的操作下可以达到90%以上的词语正确率，但在实际应用中，如电话和新闻广播等，识别率并不高。即使这样，ASR识别出来的脚本仍然对信息检索有用，这是因为检索任务只是匹配包含在音频数据中的查询词句，而不是要求一片可读性好的文章。

基于自此单元进行检索：当语音识别系统处理各方面无限制主题的大范围语音资料时，识别性能会变差，尤其当一些专业词汇（如人名、地点）不在系统词库中时。一种变通的方法是利用子词(Sub Word)索引单元，当执行查询时，用户的查询受限被分解为子单元，然后将这些单元的特征与库中预先计算好的特征进行匹配。

基于识别关键词进行检索：在无约束的语音中自动检测词或短语通常称为关键

词的发现(Spotting)。利用该技术，识别活标记出长短录音或音轨反映用户感兴趣的事情，这些标记就可以用于检索。如通过捕捉体育比赛解说词中"进球"的词语可以标记进球的内容。

基于说话人的辨认进行分割：这种技术是简单地辨别出说话人话音的差别，而不是识别出说的是什么。它在合适的环境中可以做到非常准确。利用这种技术，可以根据说话人的变化分割录音，并建立录音索引。如用这种技术检测视频或多媒体资源的声音轨迹中的说话人的变化，建立索引和确定某种类型的结构(如对话)。例如，分割和分析会议录音，分割的区段对应于不同的说话人，可以方便地直接浏览长篇的会议资料。

### 3.2.2 音频检索

音频检索是以波形声音为对象的检索，这里的音频可以是汽车发动机声、雨声、鸟叫声，也可以是语音和音乐等，这些音频都统一用声学特征来检索。

虽然 ASR 可以对语音内容给出有价值的线索，但是，还有大量其他的音频数据需要处理，从声音效果到动物叫声以及合成声音等。因此，对于一般的音频，仅仅有语音技术是不够的，使用户能从大型音频数据库中或一段长录音中找到感兴趣的音频内容是音频检索要做的事。音频数据的训练、分类和分割方便了音频数据库的浏览和查找，基于听觉特征的检索为用户提供高级的音频查询接口。这里指的音频检索就是针对广泛的声音数据的检索，分析和检索的音频可以包含语音和音乐，但是采用的是更一般性的声学特性分析方法。

① 声音训练和分类

通过训练来形成一个声音类。用户选择一些表达某类特性的声音例子(样本)，如"脚步声"。对于每个进入数据库中的声音，先计算其 N 维声学特征矢量，然后计算这些训练样本的平均矢量和协方差矩阵，这个均值和协方差就是用户训练得出的表达某类声音的类模型。

声音分类是把声音按照预定的类组合。首先计算被分类声音与以上类模型的距离，可以利用 Euclidean 或 Manhattan 距离度量，然后距离值与门限(阈值)比较，以确定是否该声音纳入或不属于比较的声音类。也有某个声音不属于任何比较的类

的情况发生，这时可以建立新的类，或纳入一个"其他"类，或归并到距离最近的类中。

② 听觉检索

听觉感知特性，如基音和音高等，可以自动提取并用于听觉感知的检索，也可以提取其他能够区分不同声音的声学特征，形成特征矢量用于查询。

例如，按时间片计算一组听觉感知特征：基音、响度、音调等。考虑到声音波形随时间的变化，最终的特征矢量将是这些特征的统计值，例如用平均值、方差和自相关值表示。这种方法适合检索和对声音效果数据进行分类,如动物声、机器声、乐器声、语音和其他自然声等。

③ 音频分割

以上方法适合单体声音的情况，如一小段电话铃声、汽车鸣笛声等。但是，一般的情况是一段录音包含许多类型的声音，由多个部分组成。更为复杂的情况是，以上各种声音可能会混在一起，如一个有背景音乐的朗诵、同声翻译等。这需要在处理单体声音之前先分割长段的音频录音。另外，还涉及到区分语音、音乐或其他声音。例如对电台新闻节目进行分割，分割出语音、静音、音乐、广告声和音乐背景上的语音等。

通过信号的声学分析并查找声音的转变点就可以实现音频的分割。转变点是度量特征突然改变的地方。转变点定义信号的区段，然后这些区段就可以作为单个的声音处理。例如，对一段音乐会的录音，可通过自动扫描找到鼓掌声音，以确定音乐片断的边界。这些技术包括：暂停段检测、说话人改变检测、男女声辨别，以及其他的声学特征。

音频是时基线性媒体。现在我们看到的典型音频播放接口是与磁带录音机相似的界面，具有停止、暂停、播放、快进、倒带等按钮。为了不丢失其中的重要东西，必须从头到尾听一遍声音文件，这样要花费很多时间，即使使用"快进"，也容易丢失重要的片断，不能满足信息技术的要求。因此，在分割的基础上，就可以结构化表示音频的内容，建立超越常规的顺序浏览界面和基于内容的音频浏览接口。

### 3.2.3 音乐检索

音乐检索是以音乐为中心的检索，利用音乐的音符和旋律等音乐特性来检索。如检索乐器、声乐作品等。

音乐是我们经常接触的媒体，像 MIDI、MP3 和各种压缩音乐制品、实时的音乐广播等。音乐检索虽然可以利用文本注释，但音乐的旋律和感受并不都是可以用语言讲得清楚的。通过在查询中出示例子，基于内容的检索技术在某种程度上可以解决这种问题。

音乐检索利用的是诸如节奏、音符、乐器特征。节奏是可度量的节拍，是音乐中一种周期特性和表示。音乐的乐谱典型地以事件形式描述，如以起始时间、持续时间和一组声学参数(基音、音高、颤音等)来描述一个音乐事件。注意到许多特征是随时间变化的，所以，我们应该用统计方法来度量音乐的特性。

人的音乐认知可以基于时间和频率模式，就像其他声音分析一样。时间结构的分析基于振幅统计，得到现代音乐中的拍子。频谱分析获得音乐和声的基本频率，可以用这些基本频率进行音乐检索。有的方法是使用直接获得的节奏特征，即假设低音乐器更适合提取节拍特征，通过归一化低音时间序列得到节奏特征矢量。

除了用示例进行音乐查询之外，用户甚至可以唱或哼出要查找的曲调。基音抽取算法把这些录音转换成音符形式的表示，然后用于对音乐数据库的查询。但是，抽取乐谱这样的属性，哪怕是极其简单的一段也是非常困难的。研究人员现在改用 MIDI 音乐数据格式解决这个问题。用户可以给出一个旋律查询，然后搜索 MIDI 文件，就可以找出相似的旋律。

# 参 考 文 献